\documentclass[11pt]{article}
\usepackage{latexsym,epsf}
\date{}
\author{Antonio O.\ Bouzas\thanks{E-mail: abouzas@fis.cinvestav.mx}
  \\\small Departamento de F\'{\i}sica
  Aplicada, CINVESTAV-IPN \\\small Apdo.\ Postal 73 ``Cordemex,''
  M\'erida 97310, Yucat\'an, M\'exico}
\title{Perturbative thresholds in the physical region}
\newcommand{\dbw}{\Delta_{\scriptscriptstyle BW} (m_{Y^*}^2)}
\begin{document}

\maketitle

\begin{abstract}
  We consider unstable-particle scattering in the context of 3-body
  processes.  We show that all partial-wave cross-sections are finite and
  positive, and the total cross-section is proportional to the
  transverse size of space in the region of on-shell particle
  exchange.  We comment on the role of loop corrections.
\end{abstract}

\section{Introduction.}
\label{sec:intro}

At tree-level, Feynman diagrams for 2-body scattering processes can
have just simple poles as singularities in the physical region.  In
particular, logarithmic thresholds only appear in higher-order
diagrams.  The singularity structure of Feynman graphs becomes
increasingly complicated as the number of external legs grows.  Thus,
we can have thresholds in the physical region of tree-level graphs if
the number of external legs is larger than four.  By physical region
we understand here real, on-shell external momenta. 

Consider, for instance, the 3-body tree-level graphs in Figs.\ (1a)
and (1b). For concreteness, we shall temporarily assume that these are
QED Feynman graphs.  In both diagrams the two upper vertices can be
viewed as a 2-body scattering process one of whose final-state
particles subsequently undergoes a further 2-body subprocess
comprising the two lower vertices.  Clearly, there is a domain of
external momenta in the physical region for these graphs where the
internal particle shared by both scattering subprocesses is
kinematically allowed to be on its mass shell. As functions of the
squared center-of-mass energy $s$, these Feynman graphs have a
branch-cut extending over the interval of values of $s$ for which the
on-shell propagation of the internal fermion is kinematically
possible.  Such singularity is not present in graph (1c), unless the
external photon lines are sufficiently off-shell.

There is another situation in which such singularity can occur in a
2-body scattering diagram.  Imagine that the wavy lines in (1a) and
(1b) are not photons but unstable massive bosons that can decay into
an $e^+ e^-$ pair.  If the momenta of the external particles connected
to these bosons in (1a) were chosen appropiately, the momentum
circulating along boson lines would lie on their resonance peak and
diagram (1a) could dominate over (1b) and other 3-body diagrams
contributing to the same amplitude.  In this kinematic region, then,
the process would be effectively 2-body, and could be represented by
graph (1c)---with the caveat, though, that (1c) is only a short-hand
notation for (1a), since unstable particles cannot be asymptotic
states.

If the external unstable particles in (1c) were long-lived enough,
such scattering process could be accesible to experiments.  We would
then be interested in computing the cross-section of a 2-body process
possessing the type of singularities discussed above.  This problem
has been posed before, in connection with several different
phenomenological situations.  For example, the process
$N^*\pi\rightarrow N^*\pi$, in which the intermediate particle is a
nucleon, was considered in \cite{prls} in relation with the so-called
``Peierl's mechanism.''  The process $\mu^+\mu^-\rightarrow WW^*$ was
considered in \cite{gnzb}, and a similar problem was discussed in the
context of an effective model Lagrangian in \cite{bzs}.  In all of
these cases the amplitudes for graphs analogous to (1c) were assumed
to have a Breit-Wigner form, in spite of the fact that the
exchanged particle is stable.  Such assumption can be heuristically
motivated by the fact that the external unstable states are on their
``mass-shell,'' which is identified with their complex pole mass
\cite{prls}. In the familiar case of resonance-formation processes in
the $s$-channel, the Breit-Wigner form of the amplitude can be
rigorously proved by Dyson resummation of the propagator followed by
Laurent expansion around the pole \cite{strt}.  As far as we know, no
proof exists in the case of processes like (1a,b,c).  Below we shall
argue that this assumption can lead to a good approximation to the
tree-level cross-section in some situations.

The singularity in 3-body diagrams like (1a) and (1b) is a
manifestation of a well-known phenomenon, namely, that a singularity
in an amplitude occurs if the interaction is unbounded in space-time.
As long as the exchanged particle in (1a,b,c) is virtual the
uncertainty principle restricts the space-time distance over which it
can propagate, but when it can be on-shell the process becomes
unbounded and the amplitude singular \cite[\S 18.5]{bjrk}.  The same
situation can occur in higher-order diagrams.  According to a theorem
by Coleman and Norton \cite{clmn} (see also \cite[\S 6.3.4]{itzk}), a
Feynman graph will possess a singularity in its physical region if and
only if the graph can be interpreted as a momentum-conserving process
in space-time with all internal lines on their mass-shell and
propagating forwards in time\footnote{Notice, however, that the notion
  of ``physical region'' used in \cite{clmn} is less restrictive than
  ours.  External lines are not required to be on-shell in \cite{clmn}.}.
The Coleman-Norton theorem is formulated for 1-particle irreducible
diagrams and valid to all orders of perturbation theory.  The
singularities in tree-level graphs such as (1a,b) have the same
physical origin.

In the particular situation considered here of graphs (1a,b,c) these
general statements can be made more quantitative.  The physical region
for $s$ in (1a,b) is of the form $0<s_{\rm th}\leq s< \infty$. The
kinematical region over which on-shell internal particles are
allowed will be limited by values $s_\pm$, and will partially or
completely overlap the physical region.  Let us assume, for
concreteness, that $s_{\rm th}<s_-<s_+$. In the region $s_-\leq s\leq
s_+$ where the internal particle can be on-shell, it can propagate
over arbitrarily long distances.  Therefore, although the largest
momentum it can have is finite, the angular momentum of the process
can be arbitrarily large.  In fact, an on-shell internal particle
will feed partial waves with essentially uniform probability, leading
to a total cross-section that behaves as,
\begin{equation}
  \label{totcros}
  \sigma = \sum_{\ell =0}^{L} \sigma^{\ell}\sim \sum_{\ell =0}^{L} 
  {\cal O}(1)\sim
  {\cal O}(L) ~~~ \mbox{as} ~~~ L\rightarrow\infty~. 
\end{equation}
Since $L\sim pb$ with $p$ the (fixed) transverse linear momentum and
$b$ the impact parameter, we conclude that the cross-section due to
on-mass-shell particle exchange diverges as $\sigma\sim b$, where $b$
is the transverse dimension of space.

On the other hand, in the region $s< s_-$ or $s> s_+$, the interaction
region is bounded.  Each partial-wave cross-section $\sigma^\ell$ is finite
and decreases rapidly with $\ell$ so that the sum converges to a
finite value $\sigma$.  As $s$ approaches $s_-$ ($s_+$) from below
(above), the intermediate particle can be closer to its mass-shell and
propagate over increasingly larger distances.  Even though for each
partial wave the transverse size of the process is fixed, since $\ell$
is fixed, in the limit $s\rightarrow s_\pm$ the longitudinal size is
unbounded and as a consequence each partial wave is singular at
$s=s_\pm$.  The singularity is logarithmic, though, and therefore
integrable, thus leading to finite results for suitably averaged
observables.  

We substantiate these assertions in the rest of this paper.  In the
next section we perform a brief kinematical analysis which serves to
establish our notation.  We also define there the cross-section for
graph (1c), which is derived from (1a).  In section 3 we show that all
partial-wave cross-sections are finite and positive and, as shown in
section 4, they result in a total cross-section wich diverges with the
transverse dimension of space as indicated above.  In section 5 we
give some final remarks, commenting in particular on the r\^ole of
loop corrections.

\section{2- and 3-body processes.}
\label{sec:body}

We consider in this section the situation described in the
Introduction, in which diagrams of the type (1a) dominate over other
diagrams contributing to the amplitude.  We will work with scalar
fields for simplicity, with couplings of the form $g \phi^\dagger\phi
X$, where $\phi$ is a light complex scalar field and $X$ a heavy real
one with $m_X > 2 m_\phi.$ We then have the graphs shown in figure 2
for the process,
\begin{equation}
  \label{proc}\hspace*{-1ex}
  \left(\phi(p_1)\overline\phi(p_2)\right) \phi(k_1) \rightarrow
  X(q_1) \phi(k_1) \rightarrow Y(q_2) \phi(k_2) \rightarrow
  \left(\phi(p_3)\overline\phi(p_4)\right) \phi(k_2)~.
\end{equation}
Here we have a different scalar $Y$ in the final state, with mass $m_Y >
2m_\phi$ and coupling $g \phi^\dagger\phi
Y$, but we shall not introduce a new coupling constant for
simplicity.  We assume that the initial stable particles are described
by plane waves and therefore have sharply defined momenta.
Furthermore, $q_1=p_1+p_2$ and $q_2=p_3+p_4$ are assumed to lie within
the resonance peak of $X$ and $Y$, respectively.

We introduce Mandelstam invariants in the usual way, by reference to
the underlying 2-body processes,
\begin{equation}
  s=(q_1+k_1)^2;~~~t=(k_2-k_1)^2;~~~u=(k_2-q_1)^2~.
\end{equation}
We then have,
\begin{eqnarray}
  \label{amps}
  {\cal M}_{3\rightarrow 3} & = & \frac{-g}{q_1^2-m_X^2+i m_X
    \Gamma_X} {\cal M}_{2\rightarrow 3} \\
  {\cal M}_{2\rightarrow 3} & = & \frac{-g}{q_2^2-m_Y^2+i m_Y
      \Gamma_Y} {\cal M}_{2\rightarrow 2} \\
    {\cal M}_{2\rightarrow 2} & = & -i g^2 \left\{
      \frac{1}{u-m_\phi^2+i 0^+} + \frac{1}{s-m_\phi^2+i 0^+}
    \right\}\label{eq:m22}
\end{eqnarray}
We retained only the pole part in the $X$ and $Y$ propagators, in view
of our assumption that their momenta are close to the peak.  Since
$q_1$ remains constant as we vary the final state momenta, we defined
an amplitude  ${\cal M}_{2\rightarrow 3}$ for the process with an
unstable $X$ particle in the initial state by factoring out the
constant propagator and production vertex.  We also defined a 2-body
amplitude ${\cal M}_{2\rightarrow 2}$ by formally applying 
Feynman rules to the 2-body process.

The $s$-variable physical region for the 3-body scattering is $s\geq
9m_\phi^2$. Since we are interested in the kinematic region where both
$X$ and $Y$ are close to their resonance peak,  we must have,
\begin{equation}
  s \geq s_{\rm th}= ({\rm max}\!(m_X,m_{Y^*}) + m_\phi)^2
\end{equation}
where we denoted $q_2^2=(p_3+p_4)^2=m_{Y^*}^2$, with $m_{Y^*}$ close
to $m_{Y}.$  We let  $q_1^2=m_X^2$ exactly for concreteness,
although it could as well have another value within $\sim\Gamma_X$
of $m_X.$

Given $q_1^2=m_X^2$ and $q_2^2=m_{Y^*}^2,$ at fixed $s$ the
squared-momentum flowing through the internal $\phi$ line is
restricted to the interval $u_-\leq u\leq u_+$ with
\begin{eqnarray}
  u_\pm & = & \frac{m_X^2+m_{Y^*}^2+2 m_\phi^2-s}{2} +
  \frac{(m_X^2-m_\phi^2)(m_{Y^*}^2-m_\phi^2)}{2s} \pm
  \nonumber\\
  &  & \pm
  \frac{1}{2s} \sqrt{(s-m_X^2+m_\phi^2)^2-4 s m_\phi^2}
  \sqrt{(s-m_{Y^*}^2+m_\phi^2)^2-4 s m_\phi^2}\label{eq:upm}~.
\end{eqnarray}
The singularity at $u=m_\phi^2$ in the amplitude falls in the physical
region if $u_-\leq m_\phi^2\leq u_+.$  From (\ref{eq:upm}) we see that
this is possible only when $s_-\leq s\leq s_+$,  where
\begin{equation}
  s_\pm = \frac{m_X^2 m_{Y^*}^2+2m_\phi^4\pm m_X m_{Y^*} \sqrt{m_X^2-4
  m_\phi^2}\sqrt{m_{Y^*}^2-4 m_\phi^2}}{2 m_\phi^2}~.
\end{equation}
Notice that
$u_+\leq m_X^2, m_{Y}^2$ as long as $s\geq s_{\rm th},$ and that
$s_-\geq s_{\rm th}$, so that the region where exchange of an on-shell
$\phi$ is kinematically allowed is entirely within the physical region
of the 2-body process, but the absorptive part of the $\phi$
propagator vanishes in that region.  
The range of variation of $t$ and
its values in the region $s_-\leq s\leq s_+$ can be obtained from
$s+t+u=m_X^2+m_{Y^*}^2+2 m_\phi^2$.

In what follows we shall assume $m_X,m_Y\gg m_\phi$ for simplicity.  In
that case we get,
\begin{equation}\label{eq:up}
  u_- = m_X^2+m_{Y^*}^2 -s~; ~~~ u_+  = \frac{m_X^2 m_{Y^*}^2}{s}~; ~~~
  u_+ - u_- = \frac{(s-m_X^2)(s-m_{Y^*}^2)}{s}
\end{equation}
instead of the more complicated expressions (\ref{eq:upm}), and
\begin{equation}
  s_-  =  m_X^2 + m_{Y^*}^2~;~~~
  s_+  =  \frac{m_X^2 m_{Y^*}^2}{m_\phi^2}\rightarrow +\infty~.
\end{equation}

Cross-sections for the
$2\rightarrow 3$ process are given by, 
\begin{equation}\label{eq:223}
  \sigma_{2\rightarrow 3}  =  \frac{1}{2 (s-m_X^2)}
  \int\!\!\frac{d^3k_2}{(2\pi)^3 2k^0_2} \frac{d^3p_3}{(2\pi)^3 2p^0_3}
  \frac{d^3p_4}{(2\pi)^3 2p^0_4}(2\pi)^4 \delta(P_{\rm
  Tot}-k_2-p_3-p_4)\, \Theta({\rm 
  cuts})\, |{\cal M}_{2\rightarrow 3}|^2
\end{equation}
$P_{\rm Tot}$ being the total 4-momentum of the process, and $\Theta({\rm
  cuts})$ a step function enforcing the cuts that define the
cross-section we are computing.  We shall only use a loose lower bound
on $m_{Y^*}^2$ as cut in this paper, for reasons to be explained
below.  $\sigma_{2\rightarrow 3}$ is conventionally defined in
(\ref{eq:223}) in terms of the flux-factor corresponding to
an initial $|X,\phi\rangle$ state.

In order to express $\sigma_{2\rightarrow 3}$ as a function of $q_2$
we make a change of variables, $q_2=p_3+p_4,~\Delta=p_4-p_3,$ and
integrate over $\Delta$ to obtain, 
\begin{equation}
  \sigma_{2\rightarrow 3}  =  \frac{g^2}{8(2\pi)^4 (s-m_X^2)}
  \int\!d^4k_2 d^4q_2 \,\delta_+\!(k_2^2) \delta(P_{\rm
  Tot}-k_2-q_2)\, 
  \frac{\Theta({\rm
    cuts})}{(q_2^2-m_Y^2)^2+m_Y^2 \Gamma_Y^2} |{\cal M}_{2\rightarrow 2}|^2~,
\end{equation}
in terms of ${\cal M}_{2\rightarrow 2}$.  Since $q_2$ is kinematically
guaranteed to be in the forward light-cone, a resolution of the
identity exists,
\begin{equation}
  1= \int_0^\infty d(m_{Y^*}^2) \delta(q_2^2-m_{Y^*}^2)
\end{equation}
which, upon insertion in $\sigma_{2\rightarrow 3}$ and integration over
$q_2$ and $k_2^0, |\mathbf k_2|,$ yields,
\begin{equation}
  \sigma_{2\rightarrow 3}  =  \frac{g^2}{(2\pi)^3 64s} \int_{-1}^{1}
  d(\cos\theta) \int_{0}^{s} d(m_{Y^*}^2)\, \frac{\Theta({\rm
  cuts})}{(m_{Y^*}^2-m_Y^2)^2+m_Y^2\Gamma_Y^2}\,
  \frac{s-m_{Y^*}^2}{s-m_X^2}\, |{\cal M}_{2\rightarrow 2}|^2
\end{equation}
where $\theta$ is the scattering angle, $\cos\theta\propto {\mathbf
  k}_1\cdot {\mathbf k}_2.$  

We can now define a 2-body cross-section by factoring out the coupling
constant and phase-space factor coming from the $Y$ decay vertex,
\begin{eqnarray}
  \sigma_{2\rightarrow 2}  & = & \frac{1}{32\pi s} \int_{-1}^{1}
  d(\cos\theta) \int_{0}^{s} d(m_{Y^*}^2)
  \Theta({\rm
    cuts}) \dbw
  \frac{s-m_{Y^*}^2}{s-m_X^2}\, |{\cal M}_{2\rightarrow
  2}|^2 \label{eq:2cr}\\
  \dbw & \equiv &
  \frac{m_Y\Gamma_Y/\pi}{(m_{Y^*}^2-m_Y^2)^2+m_Y^2\Gamma_Y^2}~.
\end{eqnarray}
This equation can be rewritten in the more suggestive form,
\begin{equation}\label{eq:suggestive}
  \sigma_{2\rightarrow 2}  =  \int_{0}^{s} d(m_{Y^*}^2)
  \Theta({\rm cuts}) \dbw\, 
  \widetilde\sigma_{2\rightarrow 2}~,
\end{equation}
where $\widetilde\sigma_{2\rightarrow 2}$ is the cross-section we
would obtain by pretending that $X$ and $Y$ are asymptotic states and
applying the Feynman rules to the 2-body process directly.  Clearly,
the same equation holds \emph{mutatis-mutandis} for the differential
cross-section $d\sigma_{2\rightarrow 2}/d(\cos\theta).$
From (\ref{eq:m22}) we immediately obtain for
$\widetilde\sigma_{2\rightarrow 2},$ 
\begin{equation}\label{eq:born}
  \widetilde\sigma_{2\rightarrow 2} = \frac{g^4}{16\pi s}
  \frac{s-m_{Y^*}^2}{s-m_X^2} \left\{ \frac{1}{s^2} + \frac{s}{m_X^2
      m_{Y^*}^2 (s_- -s)} +
  \frac{2}{(s-m_X^2)(s-m_{Y^*}^2)}\ln\left(\frac{m_X^2 m_{Y^*}^2}{s}
  \frac{1}{s_- -s}\right) 
  \right\} ~.
\end{equation}
For completeness, we also quote the form of
$\widetilde\sigma_{2\rightarrow 2}$ when we do not neglect $m_\phi$ in
comparison with $m_{X,Y}$,
\begin{equation}
  \widetilde\sigma_{2\rightarrow 2} = \frac{g^4}{16\pi s}
  \frac{|\mathbf k_2|}{|\mathbf k_1|} \left\{ \frac{1}{(s-m_\phi)^2} +
  \frac{s}{m_\phi^2}\frac{1}{(s-s_-)(s-s_+)} + \frac{1}{2}
  \frac{1}{{|\mathbf k_2| |\mathbf k_1|} (s-m_\phi^2)}
  \ln\left(\frac{u_+(s)-m_\phi^2}{u_-(s)-m_\phi^2} \right)
  \right\} ~,
\end{equation}
where $|\mathbf k_2|$ refers to its center-of-mass frame value, and
$u_\pm$ as functions of $s$ are given in (\ref{eq:upm}).  This
expression for the cross-section is valid for $s_{\rm th}\leq s < s_-$
and $s > s_+$, with $s$ ``far'' (to be quantified below) from the
singular points $s_\pm$.  In the interval between $s_\pm$ the direct
$u$-channel contribution is negative and unbounded, and the $s$-$u$ 
interference term becomes imaginary.

\section{Partial-Wave amplitudes and cross-sections.}
\label{sec:part}

The amplitude ${\cal M}_{2\rightarrow 2}$ (see (\ref{eq:m22}))
entering the 2-body cross-section defined in (\ref{eq:2cr}) can be
expanded in partial waves,
\begin{equation}
  \label{m22part}
  {\cal M}_{2\rightarrow 2} = -i g^2 \left( \frac{1}{s+i 0^+} +
  \frac{2}{u_+ - u_-} \sum_{\ell = 0}^\infty (2\ell + 1) Q_\ell (\xi +
  i0^+) P_\ell (\cos\theta)\right)~,
\end{equation}
where $u_\pm$ are defined in (\ref{eq:up}), $P_\ell$
are Legendre polynomials and $Q_\ell$ Legendre functions of the second
kind \cite{erdl,abrm}, $\theta$ is the scattering angle and, 
\begin{equation}
  \label{eq:xi}
  \xi = 1 + \frac{2 u_-}{u_+ - u_-} = 1 + \frac{2 s (m_X^2 + m_{Y^*}^2
  -s)}{(s-m_X^2) (s- m_{Y^*}^2)}~.
\end{equation}
$Q_\ell (z)$ has a logarithmic branch-cut for $z$ real, $-1 \leq z
\leq 1,$ so we see that each partial-wave amplitude has a branch-cut in the
region of on-shell $\phi$ exchange, $s_- \leq s < s_+$ ($\rightarrow+\infty$ as
$m_{\phi}/m_{X,Y}\rightarrow 0$).   Notice that in the case of a
genuine 2-body scattering the partial-wave expansion
(\ref{m22part},\ref{eq:xi}) would hold unchanged, but we would have
$\xi > 1$ for all finite $s$ in the physical region.  Only in 3-body
processes can $\xi$ go through 1, and in that case there is on-shell
particle exchange.
 
For $s_{\rm th}\leq s < s_-$ we must take the principal branch of
$Q_\ell$, \emph{i.e.}\  $Q_\ell (\xi > 1)$ real.  Therefore, as $s$
goes through $s_-,$ and $\xi$ goes through 1, the $i0^+$ prescription
results in,  
\begin{equation}\label{eq:legcut}
  Q_\ell (\xi + i 0^+) = {\mathrm Q}_\ell (\xi) - i \frac{\pi}{2}
  P_\ell (\xi)
\end{equation}
with $\mathrm Q_\ell$ a Legendre function ``on the cut''
\cite{erdl,abrm}.  The first term in (\ref{eq:legcut}) gives the
dispersive, and the second the absorptive, part of the $u$-channel
amplitude $(u+i 0^+)^{-1}.$  Therefore, they refer to ``virtual'' and
``real,'' or on-shell, particle exchange, respectively. 

We shall separate the contributions to the cross-section coming from
the regions $s<s_-$ and $s> s_-$ for convenience,
\begin{equation}\label{eq:try}
  \sigma_{2\rightarrow 2}  =  \sum_{\ell =0}^{\infty}
  \sigma_{2\rightarrow 2}^\ell~;~~~ 
  \sigma_{2\rightarrow 2}^\ell = \left(\sigma_{2\rightarrow
      2}^\ell\right)_B + \left(\sigma_{2\rightarrow 2}^\ell\right)_A,
\end{equation}
where the contributions from the regions ``below'' and ``above'' $s_-$
are found by substituting each squared partial-wave amplitude in 
expression  (\ref{eq:2cr}) for the cross-section. Thus, for $\ell>0$
we have,
\begin{eqnarray}
  \left(\sigma_{2\rightarrow 2}^{\ell>0}\right)_B & = & \frac{(2\ell
  +1)g^4 s}{4\pi (s-m_X^2)^3} 
  \int_{s-m_X^2}^s d(m_{Y^*}^2)   \, 
  \frac{\Theta({\rm cuts)}\dbw}{s-m_{Y^*}^2}
  Q_\ell^2 (\xi) \label{eq:cslv}\\
    \left(\sigma_{2\rightarrow 2}^{\ell>0}\right)_A & = & \frac{(2\ell
    +1)g^4 s}{4\pi (s-m_X^2)^3} \int_{0}^{s-m_X^2} d(m_{Y^*}^2)
  \frac{\Theta({\rm cuts)} \dbw}{s-m_{Y^*}^2}
  \left({\rm Q}_\ell^2 (\xi) + \frac{\pi^2}{4} P_\ell^2 (\xi)\right)
  \label{eq:cslr}
\end{eqnarray}
and for $\ell=0$, with the contribution from the $s$-channel diagram
in figure (2b),
\begin{eqnarray}
  \sigma_{2\rightarrow 2}^{\ell=0} & = & \frac{g^4}{16\pi s} 
  \int_0^s d(m_{Y^*}^2) \Theta({\rm cuts)}
  \dbw\, 
  \frac{s-m_{Y^*}^2}{s-m_X^2}\times\nonumber\\
  &  & \times \left\{\frac{1}{s^2}+\frac{4}{(s-m_X^2) (s-m_{Y^*}^2)}
  \left(\Theta(\xi -1) Q_0 (\xi) \rule{0pt}{12pt}  + \Theta(1-\xi)
  {\rm Q}_0 (\xi) \rule{0pt}{12pt}  \right)\right.\nonumber\\
  &  & \left. 
  +  \frac{4 s^2}{(s-m_X^2)^2 (s-m_{Y^*}^2)^2}  
  \left(\Theta(\xi -1) Q_0^2 (\xi) + \Theta(1-\xi)
    \left( {\rm Q}_0^2 (\xi) + \frac{\pi^2}{4}
  \right)\right)\right\}\label{eq:cs0v}~, 
\end{eqnarray}
where $\xi$ as a function of $m_{Y^*}^2$, $m_X^2$ and $s$ is given in
(\ref{eq:xi}), and we used (\ref{eq:up}).  We could also write
$\sigma_{2\rightarrow 2}$ as a sum of  ``virtual'' and
``real''contributions,  
obtained from (\ref{eq:try}--\ref{eq:cs0v}) by setting $P_\ell\equiv0$ or
$Q_\ell\equiv0\equiv{\rm Q}_\ell,$ respectively.

There are three potentially dangerous singularities in the integrands
of (\ref{eq:cslv}--\ref{eq:cs0v}).  In the first place, terms
involving $Q_\ell$ are integrated over $s-m_X^2\leq m_{Y^*}^2\leq s$,
so that the factor $(s-m_{Y^*}^2)^{-1}$ is singular at the upper end
of integration.  Since $\xi={\cal O}(1/(s-m_{Y^*}^2))$ as
$m_{Y^*}^2\rightarrow s$, however, that would-be pole is cancelled by
the zero of order $\ell+1$ of $Q_\ell$ at infinity.  

Secondly, (\ref{eq:cslv}--\ref{eq:cs0v}) are all singular at
$s\rightarrow (m_X^{2})^+.$  These are spurious singularities due to our
neglecting $m_\phi.$  Let us consider those integrals involving
$Q_{\ell\geq 0}$ first.  As long as $m_{Y^*}^2 >0,$ we have $\xi={\cal
  O}(1/(s-m_X^2))$  as $s\rightarrow (m_X^{2})^+$, so the singularity
is cancelled by the zero of $Q_\ell$ at infinity.  If $m_{Y^*}^2 =0,$
however,  $\xi$ remains finite as $s\rightarrow (m_X^{2})^+$ and there is
a pole in that limit.  We will introduce a cut in the
domain of integration, 
\begin{equation}
  \label{eq:cutz}
  \Theta({\rm cuts}) = \Theta(m_{Y^*}^2 - \lambda^2)
\end{equation}
with $\lambda^2>0$, so the singularity does not develop.  We will
elaborate further on the physical meaning of this kinematical cut
below.  Notice, however, that each partial-wave cross-section should be
insensitive to the precise value of $\lambda^2$, at least so long as
$\Gamma_Y \ll m_Y.$ If $\Gamma_Y \sim m_Y,$ then the very notion of an
effectively 2-body scattering comes into question and the whole
process should be dealt with, including all possible diagrams with 3
stable particles in the final state.  The cut (\ref{eq:cutz}) also
solves the problem in the other integrals not involving $Q_\ell$,
since now we must have $s\geq m_X^2 + \lambda^2 > m_X^2$ in order to
be in the region of on-shell $\phi$-exchange, and those integrals are
then strictly zero for $s<m_X^2+\lambda^2.$

Lastly, there is a branch point of $Q_\ell$ and $\rm Q_\ell$
at $\xi=1$ (or, $m_{Y^*}^2=s-m_X^2$) which is an end-point of
integration.  This is actually the only true singularity of the
integrand.  It is a logarithmic singularity, though, and therefore
integrable.  The partial-wave cross-sections
(\ref{eq:cslv}---\ref{eq:cslr}) are thus finite and positive for all
$\ell$ and $s\geq s_{\rm th}.$

These expressions for partial-wave cross-sections are to be compared with
the heuristic Ansatz \cite{prls,gnzb,bzs}, 
\begin{equation}\label{eq:ansatz}
  {\cal M}^{\rm Ansatz}_{2\rightarrow 2} = -i g^2 \left(\frac{1}{u+i
  m_Y\Gamma_Y} + \frac{1}{s+i 0^+}\right) ~,
\end{equation}
leading to partial-wave cross-sections of the form,
\begin{equation}
  \label{eq:csansatz}
  \left(\sigma_{2\rightarrow 2}^{\ell>0}\right)_{\rm Ansatz} =
  \frac{g^4}{4\pi s}\frac{1}{(u_+-u_-)^2} \frac{s-m_Y^2}{s-m_X^2}
  (2\ell +1)\left|Q_\ell
  \left( \tilde\xi+\frac{2 i m_Y \Gamma_Y}{u_+-u_-}\right)\right|^2~,
\end{equation}
where $\tilde\xi$ is given by (\ref{eq:xi}) with $m_{Y^*}=m_Y$.  Both
(\ref{eq:cslv}---\ref{eq:cslr}) and (\ref{eq:csansatz}) approach
(\ref{eq:born}) in the narrow resonance limit $\Gamma_Y/m_Y\rightarrow
0$, and depend on $\Gamma_Y$ as $\ln^2(\Gamma_Y)$ at the maximum at
$s=s_-$.  Therefore, we expect (\ref{eq:cslv}---\ref{eq:cslr}) and
(\ref{eq:csansatz}) to be close to each other for moderate values of
$\Gamma_Y/m_Y$ and $\ell$.  We notice, however, that
(\ref{eq:csansatz}) leads to a finite total cross-section over the
whole kinematical range $s\geq s_{\rm th}$, so that it must have a
different $\ell$-dependence for large values $\ell\begin{array}{c}
  >\\[-1.5ex] \sim\end{array} m_Y/\Gamma_Y$.

Before turning to total cross-sections, we would like to remark that
equations (\ref{eq:cslv}---\ref{eq:cslr}) for $\sigma_{2\rightarrow
  2}^\ell$ were derived in the idealized assumption that the
initial-$X$ momentum is sharply defined and, in particular, its
squared-momentum is precisely known.  Let us assume now that $q_1$ in
figure (2) is statistically distributed, with a flat
probability density $\cal F$ over a range of values centered around
$q_1^2=m_X^2$.  The squared amplitude in this case takes the form of
an incoherent sum, schematically,
\begin{equation}
  \int d(m_{X^*}^2)\, {\cal F}(m_{X^*}^2)
  \Delta{\scriptscriptstyle BW} (m_{X^*}^2) \int d(m_{Y^*}^2)\, \dbw
  |{\cal M}_{2\rightarrow 2}|^2~.
\end{equation}
If ${\cal F}(m_{X^*}^2)$ is very narrow, or if $\Gamma_X/m_X\rightarrow
0$, we recover (\ref{eq:cslv}---\ref{eq:cs0v}).  If, instead,
$\Gamma_Y/m_Y\ll  \Gamma_X/m_X$, then we can replace $\dbw\sim \delta
(m_{Y^*}^2-m_Y^2)$ and we have, 
\begin{equation}\label{eq:otrans}
  \int d\!(m_{X^*}^2)\, {\cal F}(m_{X^*}^2)
  \Delta{\scriptscriptstyle BW} (m_{X^*}^2) |{\cal M}_{2\rightarrow 2}|^2~.
\end{equation}
If $\Delta_{\scriptscriptstyle BW} (m_{X^*}^2)$ is narrower than
${\cal F}(m_{X^*}^2)$, the latter is essentially constant over the
range of integration so it can be taken outside the integral and we
end up with an expression analogous to
(\ref{eq:cslv}---\ref{eq:cs0v}), but with  $\Delta_{\scriptscriptstyle
  BW} (m_{X^*}^2)$ in place of $\dbw$.  In this case, the correct
Ansatz is (\ref{eq:ansatz}) with $m_X\Gamma_X$ instead of
$m_Y\Gamma_Y$.   A similar result obtains if we consider wave-packets
as initial state rather than an incoherent superposition.  In that
case the methods of \cite{mlnk} must be applied, but we shall not go into
the details here.
 
\section{Total cross-sections.}
\label{sec:tcs}

A direct application of Feynman rules in the Born approximation to the
graphs shown in figure 2 leads to the total cross-section
(\ref{eq:born}), which is singular as $s\rightarrow s_-$ from below
and negative for $s > s_-.$ As we have seen in the previous section,
all partial-wave cross-sections $\sigma^\ell_{2\rightarrow 2}$ are finite
and positive.  If we try to compute $\sigma_{2\rightarrow 2}$ as the sum
of partial-wave cross-sections (\ref{eq:try}), however, we are
obviously led back to 
(\ref{eq:suggestive},\ref{eq:born}) as can be explicitly checked.
% by using the
%fundamental integral representation for $Q_\ell$ \cite{erdl,abrm} and
%the orthogonality of $P_\ell.$
This fact tells us that the
singularities in the total cross-section are due to the bad
convergence properties of the sum in (\ref{eq:try}) or, equivalently,
%to the bad convergence properties 
of $\sigma^\ell_{2\rightarrow 2}$ as $\ell\rightarrow\infty.$ The
physical reasons for this bad convergence were discussed in the
Introduction.  In this section we will quantitatively study the
convergence properties of $\sigma^\ell_{2\rightarrow 2}.$  In the next
two subsections we present some simple but somewhat technical results
which we discuss below in subsection 4.3.

\subsection{The region of on-shell exchange}

Let us first consider the term in (\ref{eq:cslr}) that involves $P_\ell.$ It can
be rewritten as, 
\begin{equation}
  \label{eq:pphi}
  \int_{-1}^{1} d\xi \varphi(\xi) P_\ell^2 (\xi)~,
\end{equation}
where $\varphi$ represents the combination of factors in front of $P_\ell$ in
the integrand of (\ref{eq:cslr}).  In particular, $\varphi$ contains a factor
$\Theta({\rm cuts})$.  As we will now see, the dependence of (\ref{eq:pphi}) on
$\ell$ is not sensitive to the details of the cut, such as the precise value of
$\lambda^2$ or the form of $\Theta$. We will take $\Theta$ not as a
step-function, but as some smooth approximation to it.  In this way we can
consider $\varphi$ to be continuous, non-negative and not everywhere-vanishing
on the interval of integration.

We will now show that (\ref{eq:pphi}) is ${\cal O}(1/\ell)$ as
$\ell\rightarrow\infty$.  This may seem obvious in view of the
normalization of $P_\ell$, but the key-point here is that as long as
$\varphi$ is positive and continuous the convergence will not be
faster, independently of the particular form of $\varphi$.  Since
$\varphi$ is continuous over a compact interval, it can be uniformly
approximated by a polynomial to any desired accuracy, due to a theorem
of Weierstrass \cite{zygm}.  Therefore, we need only consider the
case of polynomial $\varphi$.  Since it is independent of $\ell$, 
we may integrate $\varphi$ term by term in (\ref{eq:pphi}) and study the
convergence of each term separately.  Furthermore, since $P_\ell^2$ is
even, we need only consider even powers of $\xi$.

Thus, we have to see that,
\begin{equation}
  \label{hypo}
  \int_{-1}^{1} dx \left( x^n P_\ell (x)\right)^2 = {\cal O}
  \left(\frac{1}{\ell}\right) ~~~~ n=0,1,2,\ldots
\end{equation}
The case $n=0$ is just the normalization of $P_\ell$.  Let us consider
$n=1$,  
\begin{equation}\label{eq:recu}
  x P_\ell (x) = \frac{\ell +1}{2\ell +1} P_{\ell+1}(x) +
  \frac{\ell}{2\ell+1} P_{\ell -1}(x)
\end{equation}
\cite{erdl,abrm}.  We notice that the coefficients on the right-hand
side of (\ref{eq:recu}) are positive, and that they add up to 1,
\emph{i.e.,} $x P_\ell (x)$ is a convex combination of $P_{\ell\pm
  1}(x)$.  It easily follows by induction that a similar relation
holds for all $n$ and $\ell$,
\begin{eqnarray}
  \label{eq:convex}
  x^n P_\ell (x) = \sum_{k=-n}^{n} {\hspace*{-1ex}}^\prime a_k(n,\ell)
  P_{\ell+k} (x)~;~~~ 
  a_k(n,\ell)> 0 ~;~~~   \sum_{k=-n}^n {\hspace*{-1ex}}^\prime a_k(n,\ell) = 1~,
\end{eqnarray}
where the prime indicates that the sum runs only over $k$ with the same parity
as $n$.  Since the second of (\ref{eq:convex}) is true for
all $\ell$ and $n$, we must have,
\begin{equation}
  \label{eq:limconv}
  \lim_{\ell\rightarrow\infty} a_k(n,\ell) =
  a_k(n)~~~\mbox{with}~~~0\leq a_k(n) \leq 1
\end{equation}
and not all $a_k(n)=0$.  Thus, we can write,
\begin{equation}
  \int_{-1}^{1}\!\!dx \left( x^n P_\ell (x)\right)^2  =  \int_{-1}^1\!\!dx
  \!\!\!\sum_{k,j=-n}^n \!\!a_k(n,\ell) a_j(n,\ell) P_{\ell+k}(x)
  P_{\ell+j}(x)
   =\!\!\sum_{k=-n}^n \!\! a_k^2(n,\ell) \frac{2}{2(\ell+k)+1}
\end{equation}
and, since $n$ and $k$ are independent of $\ell$, we can take the
limit $\ell\rightarrow\infty$ to obtain,
\begin{equation}
  (2\ell +1) \int_{-1}^{1} (x^n P_\ell (x))^2 \longrightarrow
  2 \sum_{k=-n}^n a_k^2(n) ~. 
\end{equation}

The same argument goes through for the term in (\ref{eq:cslr}) involving ${\rm
Q}_\ell$, because  recursion relation (\ref{eq:recu}) is valid if we
substitute ${\rm Q}_\ell$ for $P_\ell$ there,  and because \cite{erdl}
\begin{equation}
  \int_{-1}^1 dx\,{\rm Q}_\ell^2 (x)  = {\cal
  O}\left(\frac{1}{\ell}\right)
~;~~~
  \int_{-1}^1 dx\,{\rm Q}_{\ell+n} {\rm Q}_{\ell+m}  =  {\cal
  O}\left(\frac{1}{\ell^2}\right) ~~~ (m\neq n) 
\end{equation}
as $\ell\rightarrow\infty$.  Therefore, $(\sigma^\ell_{2\rightarrow 2})_A$
is a finite constant in the limit $\ell\rightarrow\infty$ and we have,
\begin{equation}\label{eq:sigAL}
  \left(\sigma_{2\rightarrow 2}\right)_A =
  \lim_{L\rightarrow\infty} \sum_{\ell=0}^L
  \left(\sigma^\ell_{2\rightarrow 2}\right)_A  = {\cal O}(L)~,
\end{equation}
as claimed in the Introduction.

\subsection{The region of off-shell exchange}

We now turn to integral (\ref{eq:cslv}), which involves $Q_\ell$ and can be
written in the form,
\begin{equation}\label{eq:form}
  \int_{1}^\infty\! d\xi\, \varphi (\xi) \left[(s+m_X^2) + (s-m_X^2)\xi\right]
  Q_\ell^2 (\xi)~,
\end{equation}
where $\varphi$ represents the combination of factors in the integrand in
(\ref{eq:cslv}), but we have explicitly extracted a term linear in $\xi$ coming
from the denominator $(s-m_{Y*}^2)$ in (\ref{eq:cslv}) which diverges at the
upper limit of integration.  We shall assume that $\varphi (\xi)$ is continuous
and bounded in the interval of integration and that it is positive, and non-zero
for $1\leq\xi\leq a$, for some finite $a>1$.  The integrand in (\ref{eq:cslv})
satisfies all of these conditions once the singular term $(s-m_{Y*}^2)^{-1}$ is
factored.  First, we show that,
\begin{equation}\label{eq:ineq}
  \int_{a}^\infty \!d\xi\, \varphi (\xi) [(s+m_X^2)+(s-m_X^2)\xi]
  Q_\ell^2 (\xi) \leq 
  \frac{\rm cst}{a^{(2\ell-1)} \ell}; ~~~a>1~.
\end{equation}
Notice that since $a>1$ the right-hand side of this equation is
negligibly small for large $\ell$, and that better bounds can be
found.  We can write \cite{erdl}, for $\xi>1$,
\begin{eqnarray}
  Q_\ell (\xi) & = &\sqrt{\frac{\pi}{2}}
  \frac{\left(\xi-\sqrt{\xi^2-1}\right)^{\ell+\frac{1}{2}}}{(\xi^2-1)^\frac{1}{4}}
  \frac{\Gamma(\ell+1)}{\Gamma(\ell+\frac{3}{2})}\, {}_2\!F_1
  \left(\frac{1}{2},
  \frac{1}{2},\ell+\frac{3}{2},\frac{-\xi+\sqrt{\xi^2-1}}{2
    \sqrt{\xi^2-1}}\right) \nonumber\\
  &\leq & \sqrt{\frac{\pi}{2}}
  \frac{\left(\xi-\sqrt{\xi^2-1}\right)^{\ell+\frac{1}{2}}}{(\xi^2-1)^\frac{1}{4}}
  \frac{\Gamma(\ell+1)}{\Gamma(\ell+\frac{3}{2})}
  \leq \sqrt{\frac{\pi}{2}} \frac{1}{\xi^{\ell+\frac{1}{2}}}
  \frac{1}{(\xi^2-1)^\frac{1}{4}}
  \frac{\Gamma(\ell+1)}{\Gamma(\ell+\frac{3}{2})}~.
\end{eqnarray}
Substitution of this inequality in the left-hand side of
(\ref{eq:ineq}) leads to the desired result, since $\varphi$ is
bounded.  

Next, we show that,
\begin{equation}
  \int_1^a\!d\xi\, \varphi(\xi) [(s+m_X^2)+(s-m_X^2)\xi] Q_\ell^2(\xi) = {\cal
  O}\left(\frac{1}{\ell^2}\right) ~~~ \mbox{as}~ \ell \rightarrow\infty~.
\end{equation}
We can follow the same procedure as in the previous section, the only
difference being that now odd powers of $\xi$ must also be
considered.  We have \cite{erdl,abrm},
\begin{equation}
  \int_1^a\!d\xi\, Q_\ell^2(\xi) = \int_1^\infty\!d\xi\, Q_\ell^2(\xi)
  -\int_a^\infty\!d\xi\, Q_\ell^2(\xi) =\frac{\psi^\prime
  (\ell+1)}{2\ell+1} - o\left(\frac{1}{a^{2\ell -1}\ell}\right)
  \rightarrow \frac{1}{\ell^2} ~,
\end{equation}
where $\psi$ is a digamma function \cite{erdl,abrm}.  The case of even
powers then follows from here since $Q_\ell$ 
satisfies the same recursion relation (\ref{eq:recu}) as $P_\ell$, and
since, 
\begin{eqnarray}
  \int_1^a\!d\xi\, Q_{\ell+n}(\xi) Q_{\ell+m}(\xi) & = &
  \int_1^\infty\!d\xi\, Q_{\ell+n}(\xi) Q_{\ell+m}(\xi) -
  \int_a^\infty\!d\xi\, Q_{\ell+n}(\xi) Q_{\ell+m}(\xi)\nonumber\\
  &=& \frac{\psi(\ell+n+1)-\psi(\ell+m+1)}{(n-m) (2\ell+n+m+1)} -
  o\left(\frac{1}{a^{2\ell-1}\ell}\right) \rightarrow \frac{1}{\ell^2}~.
\end{eqnarray}
We can then apply the same arguments as above.  For odd powers we
have, $\xi^{2n} Q_\ell^2(\xi)\leq\xi^{2n+1}Q_\ell^2(\xi)\leq\xi^{2n+2}
Q_\ell^2(\xi)$ for $1\leq\xi\leq a$ and hence the same result must
hold.  Finally, 
\begin{equation}\label{eq:csBL}
  (2\ell+1) \int_1^\infty\!d\xi\, \varphi(\xi)
  [(s+m_X^2)+(s-m_X^2)\xi] Q_\ell^2
  (\xi)\stackrel{\ell\rightarrow\infty}{\longrightarrow} \frac{\rm
  cst}{\ell} ~,
\end{equation}
as asserted above.

\subsection{Remarks}

The preceding results show how the cross-section behaves as a function
of $s$, and the influence of the kinematical cut $\lambda^2$.  Let us
take $\lambda^2 < m_Y^2$, and consider $(\sigma_{2\rightarrow
  2}^\ell)_A$.  For $s<m_X^2+\lambda^2$, $(\sigma_{2\rightarrow
  2}^\ell)_A=0.$ When $m_X^2+\lambda^2< s<m_X^2+m_Y^2$, we have
$(\sigma_{2\rightarrow 2}^\ell)_A>0$ and, as shown in
(\ref{eq:sigAL}), $(\sigma_{2\rightarrow 2})_A$ diverges.  Both
$(\sigma_{2\rightarrow 2}^\ell)_A$ and $(\sigma_{2\rightarrow 2})_A$
are of ${\cal O}(g^6)$, and therefore negligible in the 2-body
approximation, because the integral in $(\sigma_{2\rightarrow
  2}^\ell)_A$ does not include the peak of $\dbw$.  It is only when
$s-m_X^2\begin{array}{c} > \\[-2ex] \sim \end{array} m_Y^2$ that
$(\sigma_{2\rightarrow 2}^\ell)_A$, $(\sigma_{2\rightarrow 2})_A$ are
actually non-vanishing in the 2-body approximation.  Thus, in this
case, the precise value of $\lambda^2$ is irrelevant as long as it
lies below the peak of $\Delta_{BW}$.

The cut does influence the dependence of $(\sigma_{2\rightarrow
  2}^\ell)_B$ on $s$.  In the region $s<m_X^2+\lambda^2$, the branch
point at $m_{Y^*}^2=s-m_X^2$ (or, $\xi=1$) is excluded by $\Theta({\rm
  cuts})$ from the integration region in (\ref{eq:cslv}).  Thus,
(\ref{eq:ineq}) applies, implying that $(\sigma_{2\rightarrow
  2}^\ell)_B$ converges to zero fast enough to make
$(\sigma_{2\rightarrow 2})_B$ finite.  We are here essentially in
the same situation as in (\ref{eq:born}) with $s<s_-$.

When $s>m_X^2+\lambda^2$, on the other hand, the range of integration
in (\ref{eq:cslv}) extends all the way down to $m_{Y^*}^2=s-m_X^2$ (or, $\xi =1$
in (\ref{eq:form})) and (\ref{eq:csBL}) holds, leading to a cross-section
$(\sigma_{2\rightarrow 2})_B$ diverging with $L$ as $\log(L)$.  For
$s\begin{array}{c} > \\[-2ex] \sim \end{array} m_X^2+m_Y^2$, finally,
the peak of $\dbw$ leaves the integration region and
$(\sigma_{2\rightarrow 2}^\ell)_B$, $(\sigma_{2\rightarrow 2})_B$
become ${\cal O}(g^6)$ and therefore negligible.

We see, then, that the value of $\lambda^2$ determines the point
$s=m_X^2+\lambda^2$ at which $(\sigma_{2\rightarrow 2})_B$ is
singular, \emph{i.e.,} it is the lowest value of $m_{Y^*}^2$ at which
we start considering $Y$ to be ``on-shell.'' It is therefore
inherently ambiguous.  In the case of (\ref{eq:born}), this value is taken as
$m_Y^2$.  When we take into account the finite width of $Y$,
$\lambda^2$ should be set to $\lambda^2=m_Y^2- \alpha m_Y\Gamma_Y$
with $\alpha\sim 1$--2.

It is worth-while to stress here that the
need for a cut $\lambda^2$ is independent of the kinematical
singularity mentioned in the paragraph preceding eq.\ (\ref{eq:cutz}).
Even if we set 
$m_\phi>0$, a cut in $m_{Y^*}^2$ is needed.  Notice, however, that if
$m_\phi>0$ we would also need an upper cut $\Theta(\lambda^{\prime
  2}-m_{Y^*}^2)$, where now $\lambda^{\prime 2}=\lambda^2=m_Y^2+ \alpha
m_Y\Gamma_Y$, $\alpha\sim 1$--2.  Since for $m_\phi=0$ the region of
on-shell particle exchange extends to infinity, no upper kinematical
cut is needed (or possible)---actually, $\dbw$ does the job
as described in the previous paragraph.  

\section{Final Remarks}

In the foregoing sections we treated the problem of unstable-particle
scattering in the context of 3-body reactions.  We showed how the
cross-sections can be defined in sections 2 and 3.  The spatial
singularities of the total cross-section were characterized in section
4.  The result found there for the total cross-section in the region
of on-shell exchange, $\sigma_{2\rightarrow 2}={\cal O}(L)$, agrees
with the conclusions of \cite{mlnk}.  Indeed, an initial state
consisting of wave-packets has an angular-momentum cut-off $L_{\rm
  max}$.  Therefore, we must have in that case $\sigma_{2\rightarrow
  2}\sim L_{\rm max}\sim b$, with $b$ the largest available impact
parameter, and then $\sigma_{2\rightarrow 2}$ grows linearly with the
transverse size of the wave-packets.  In the general case of
asymmetric wave-packets, of different radii in each beam,
$\sigma_{2\rightarrow 2}$ must be proportional to the largest one.  In
the region of off-shell exchange, $\sigma_{2\rightarrow 2}\sim \log L$
as we approach the singular point.

The results of the previous sections were obtained at tree-level, with
the exception of the Breit-Wigner form used for the propagators of
unstable particles near their resonance peak.  It is for this reason
that we had to introduce an arbitrary mass cut-off $\lambda$ that
separates the resonance peak, where the amplitude is ${\cal O}(g^4)$,
from the tail where it is  ${\cal O}(g^6)$.  This cut $\lambda$ is
inherently ambiguous, $\lambda^2=m^2-\alpha m\Gamma$ with $\alpha\sim
1$--2, and should not be necessary in a more complete treatment
including loop corrections.  We notice also that partial-wave
cross-sections $\sigma_{2\rightarrow 2}^\ell$ are of order $\sim g^4
\log (\Gamma^2/m^2)/(m^2 s^2)$ at their maximum.  They are therefore far
from saturating unitarity bounds as happens in the case of $s$-channel
resonance formation, where $\sigma^{\ell=0}_{2\rightarrow 2} \sim {\cal
  O}(1)$ in the coupling constant.

Let us consider one more time the Feynman graph for the process,
redrawn in figure 3a for convenience.  For $s_-<s<s_+$, the
intermediate state indicated by the dashed line can be formed by
particles 2, 3, 4 propagating on their mass-shell and forward in time.
Given that particles 2 and 3 do interact, we ask what is the
probability that these particles will propagate over a long distance
before the interaction takes place.  The heuristic Ansatz
(\ref{eq:ansatz}) implies that this probability decays exponentially
with rate $1/\Gamma_5$, the characteristic resonance formation time
for particle 5.

We have shown that at tree level there is no characteristic
length-scale governing the process, thus leading to
$\sigma_{2\rightarrow 2}={\cal O}(L)$ as $L\rightarrow\infty$.  That
result is most likely not changed by loop corrections when 1 and 5 in
figure 3a are stable virtual particles.  Whether the tree-level result
is altered by loop corrections when unstable particles are involved
and, in particular, whether loop corrections lead to (\ref{eq:ansatz})
is currently under study.  It is worth remarking here that equation
(\ref{eq:otrans}) suggests that Ansatz (\ref{eq:ansatz}) may not be the
complete answer when more than one species of unstable particle are
involved. 

Self-energy loop insertions such as graph 3b cannot change the basic
conclusions reached at tree-level, although they lead to an amplitude
with a stronger singularity $\sim 1/u^2$, than that found at
tree-level, $\sim 1/u$.  Resummation of these corrections, however,
will only lead to mass and wave-function renormalization, since
particle 3 is stable and its self-energy cannot have an absorptive
part at the pole.  The opposite is true for vertex correction 3c,
which will have a non-vanishing absorptive part when particle 3 is
on-shell, but depends on $u$ as $\sim 1/u$.  As long as the coupling
constant is small, this absorptive part should be a higher-order
correction to the tree-level amplitude.  We are then left with
1-particle irreducible graphs like 3d and 3e.  Regarding those, we
shall limit ourselves to point out that the unitary cut depicted in
figure 3f leads directly to equation (\ref{eq:suggestive}) for the
cross-section.  Notice that the extra loop in 3f is necessary
\cite{vltm}, since the propagator of an unstable particle does not
have a pole in the physical region.

\section*{Acknowledgements}

I am indebted to Profs.\  V.\ Gupta and C.A.\ Garc\'{\i}a
Canal for their criticism of earlier versions of this paper and
for suggesting many improvements.  I also benefitted from a remarkably
accurate referee's report.

\noindent I would like to thank Profs.\ J.M.\ Cornwall for pointing out ref.\
\cite{clmn} to me, and G.\ L\'opez Castro for bringing ref.\
\cite{gnzb} to my attention and for several discussions.

\noindent I would also like to thank Profs.\ A.\ Zepeda
Dom\'{\i}nguez, M.A.\ P\'erez Ang\'on, A.\ Garc\'{\i}a Gonz\'alez and R.\
Huerta Quintanilla for help and support during these years at CINVESTAV.

\noindent This work has been partially supported by CONACYT through
SNI and Research Project 0247P.

\section*{Figure Captions}

\begin{description}
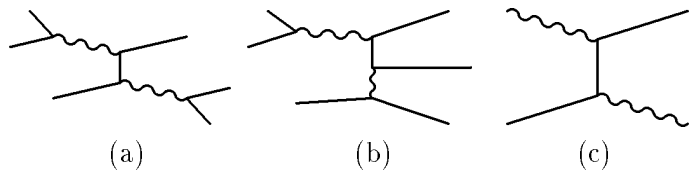
\item[Figure 1] Generic Feynman diagrams for 3- and 2-body scattering
      processes.
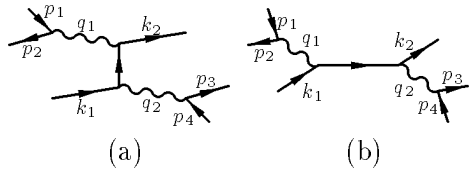
\item[Figure 2] 3-body scattering graphs in the kinematical region where
      unstable internal particles are close to their pole mass.
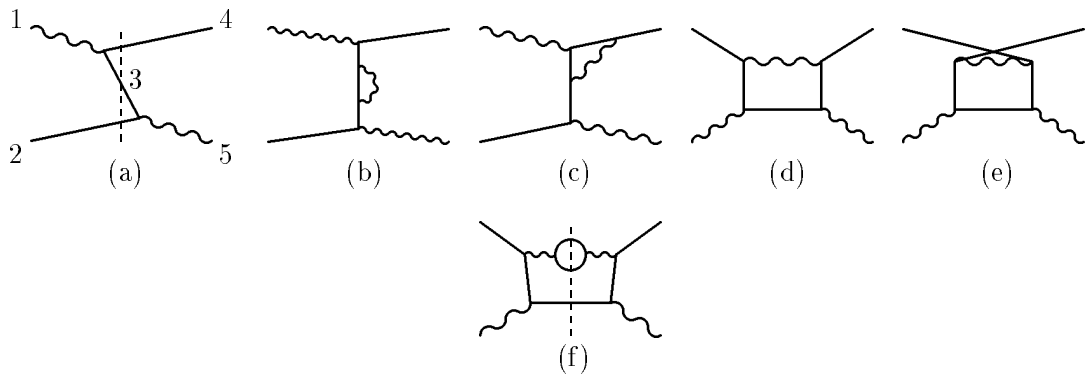
\item[Figure 3] (a)The graph in figure 2a is redrawn here omitting the
      external legs labelled $p_{1\mbox{--}4}$.  Particles 1 and 5
      correspond to $X,Y$ and 2,3,4 to $\phi,$ resp.,  in fig.\ 2a.
      (b)--(e) Some one-loop corrections to the process depicted
      in (a). (f) A unitary cut in (d).
\end{description}

\end{document}